\documentclass[letterpaper,onecolumn,12pt]{IEEEtran}
\ifCLASSINFOpdf
\else
\fi
\usepackage{flushend}
\usepackage{stfloats}
\usepackage{fancyhdr}
\usepackage{mathrsfs}
\usepackage{graphicx}
\usepackage{color}
\usepackage{placeins}
\usepackage{float}
\usepackage{tabularx,colortbl}
\usepackage{amsmath}
\usepackage{amsfonts}
\usepackage{amssymb}
\usepackage{amsthm}
\usepackage{bm}
\usepackage[numbers,sort&compress]{natbib}
\usepackage{setspace}
\usepackage{graphicx}
\usepackage{subfigure}
\begin{document}
%
\title{Cascaded Channel Estimation for IRS-assisted Mmwave Multi-antenna with Quantized Beamforming}
\author{\IEEEauthorblockN{Wenhui~Zhang, Jindan~Xu\IEEEauthorrefmark{0}, \IEEEmembership{Student Member,~IEEE}, Wei~Xu\IEEEauthorrefmark{0}, \IEEEmembership{Senior Member,~IEEE}, \\
Derrick~Wing~Kwan~Ng\IEEEauthorrefmark{0}, \IEEEmembership{Senior Member,~IEEE}}, and Huan~Sun\IEEEauthorrefmark{0}, \IEEEmembership{Member,~IEEE}\\
\vspace{-1cm}
\thanks{W. Zhang, J. Xu, and W. Xu are with the National Mobile Communications Research Laboratory (NCRL), Southeast University, Nanjing 210096, China (\{whzhang, jdxu, wxu\}@seu.edu.cn).

D. W. K. Ng is with the School of Electrical Engineering and Telecommunications, University of New South Wales, Sydney, NSW 2052, Australia (w.k.ng@unsw.edu.au).

H. Sun is with Huawei Technologies Shanghai R\&D Center, Shanghai 201206, China (sunhuan11@huawei.com).}
}


%


\maketitle
\thispagestyle{fancy}
\renewcommand{\headrulewidth}{0pt}
\pagestyle{fancy}
\cfoot{}
\rhead{\thepage}

\newtheorem{mylemma}{Lemma}
\newtheorem{mytheorem}{Theorem}
\newtheorem{mypro}{Proposition}
\vspace{-0.8cm}
\begin{abstract}
In this letter, we optimize the channel estimator of the cascaded channel in an intelligent reflecting surface (IRS)-assisted millimeter wave (mmWave) multi-antenna system. In this system, the receiver is equipped with a hybrid architecture adopting quantized beamforming. Different from traditional multiple-input multiple-output (MIMO) systems, the design of channel estimation is challenging since the IRS is usually a passive array with limited signal processing capability. We derive the optimized channel estimator in a closed form by reformulating the problem of cascaded channel estimation in this system, leveraging the typical mean-squared error (MSE) criterion. Considering the presence of possible channel sparsity in mmWave channels, we generalize the proposed method by exploiting the channel sparsity for further performance enhancement and computational complexity reduction. Simulation results verify that the proposed estimator significantly outperforms the existing ones.
\end{abstract}


%
\IEEEpeerreviewmaketitle

\section{Introduction}

\IEEEPARstart{M}{illimeter} wave (mmWave) massive multiple-input multiple-output (MIMO) promises an order of magnitude increase in spectral efficiency of wireless communication \cite{MIMO5G}. However, massive MIMO is characterized by an enormous antenna array, requiring a large number of radio-frequency (RF) chains, which is costly and energy-consuming. In particular, an analog-to-digital converter (ADC) is one of the basic infrastructures of a RF chain, significantly contributing to the extra cost and energy consumption.

To address this issue, one can resort to a hybrid analog-and-digital architecture with a limited number of RF chains \cite{AyachSPAWC}, where low-precision ADCs can also be exploited to further reduce power consumption \cite{JXuADC}. To alleviate the high cost and power consumption of massive MIMO, intelligent reflecting surface (IRS) consisting of a mass of passive reflecting elements emerges as a complementary technology, whose superiority in flexibly manuipulating electromagnetic wave is evident \cite{B5G}. In particular, by altering the phases of reflected signals, IRS enables energy focusing and energy nulling at desired locations via beamforming. There are numerous potential use cases of IRS and multiple-input single-output (MISO)/single-input multiple-output (SIMO) which cover a wide range of practical scenarios, e.g., enhancing cell-edge coverage against blockage and enabling cost-and-energy efficient communication especially at mmWave band \cite{QWuTowards}.

In spite of the enormous merits of aforementioned techniques, there are a number of challenges of channel estimation in practice. In massive MIMO adopting hybrid architecture, it is difficult to accurately estimate a high-dimensional channel matrix from a low-dimensional observation using a limited number of RF chains. As a remedy, a compressive sensing (CS)-based channel estimator was proposed by exploiting the channel sparsity \cite{alkhateeb2014channel}. It was then extended in \cite{YuWangWCL} to a hybrid MIMO system with low-precision ADCs, incorporating orthogonal matching pursuit (OMP) for sparse channel recovery.

Furthermore, the problem of channel estimation becomes more challenging when IRS is deployed in communication networks. Firstly, the signaling overhead of channel estimation increases dramatically since the number of reflecting elements of an IRS is usually large and the inherent two-hop channels result in a high dimensionality compared to existing networks \cite{BzhengIRS}. To be specific, the composite cascaded channel changes dynamically with the reflection matrix adopted at the IRS even though the physical channel remains static. Secondly, since IRSs are mostly passive which cannot transmit and receive pilot signals, it is less tractable to construct separate channel estimation as in conventional MIMO networks. For channel estimation in passive IRS-assisted systems, a typical solution is the ``on-off'' scheme, which turns on only one IRS element in each sub-phase in order to realize separate channel estimation at the cost of huge signaling overhead \cite{QIRSMISO}. In \cite{ZCascaded}, a bilinear generalized approximate message passing (BiG-AMP)-assisted algorithm was proposed by exploiting matrix decomposition on the cascaded channel, which is equivalent to a random ``on-off'' scheme. Moreover, to reduce the pilot consumption, a three-phase channel estimation protocol was proposed by fully utilizing the common channel between the IRS and the base station (BS) \cite{Cuimuser}. Following the same philosophy, the authors in \cite{DaiTwo} proposed a two-timescale channel estimation framework adopting a dual-link pilot transmission scheme, which was further extended in \cite{Anchor} by introducing anchor nodes to assist the estimation of the common BS-IRS channel with reduced pilot consumption.

Apparently, when considering the application of IRS in the popular setup, i.e., hybrid massive MIMO, the problem of channel estimation is even more problematic. In particular, due to the hybrid architecture, there are only limited observations for estimating a high-dimensional channel matrix. However, the dimension of the cascaded channel matrix of an IRS-assisted system increases rapidly due to the numerous number of IRS elements, requiring more observations for channel estimation. To the best of our knowledge, it is still an open problem to estimate the cascaded MIMO channel for IRS-assisted hybrid MIMO, especially with low-precision ADCs. In this letter, we propose a cascaded channel estimation method in a hybrid structure mmWave multi-antenna assisted by an IRS. A uniform planar array (UPA) is deployed at a BS while low-precision ADCs are deployed at the receiver for the sake of low cost. We derive a closed-form expression of the optimal linear channel estimator which alleviates the impact of the distortion caused by nonlinear quantization of the low-precision ADCs. Furthermore, if the channel sparsity is known as \emph{a prior}, we show that the proposed estimator can exploit this information and be further enhanced with lifted performance and reduced complexity.

\emph{Notations}: Throughout this paper, ${(\cdot)}^{\rm H}$, ${(\cdot)}^{\rm T}$, ${(\cdot)}^*$, ${(\cdot)}^{\dagger}$, $\mathbb{R}(\mathbb{C})$, $\otimes$, $\mathbb{E}\{\cdot\}$, ${\rm tr}(\mathbf{X})$, and $\|\mathbf{X}\|_{\rm 2}$ denote the conjugate transpose, transpose, conjugate, Moore-Penrose inverse, space of real (complex) numbers, Kronecker product operator, expectation operator, trace, and Frobenius norm, respectively. The $k$th entry of vector $\boldsymbol x$ and the $(i,j)$th element of matrix $\mathbf{X}$ are represented by $[\boldsymbol x]_{k}$ and $[\mathbf{X}]_{ij}$, respectively. We adopt ceiling $\lceil a\rceil$ to return the smallest integer no smaller than $a \in \mathbb{R}$. Operators ${\rm vec}(\mathbf{X})$ and ${\rm mat}(\boldsymbol x)$ imply that $\boldsymbol x={\rm vec}(\mathbf{X})$ is the column-stacked form of $\mathbf{X}$ and $\mathbf{X}={\rm mat}(\boldsymbol x)$ for $\mathbf{X}\in\mathbb{C}^{M \times N}$, $\boldsymbol x\in \mathbb{C}^{MN \times 1}$, respectively. $\mathbf{1}\in \mathbb{R}^{N}$ is the $N \times 1$ vector of all ones. $\mathcal{CN}(0,1)$ represents the distribution of a circularly symmetric complex Gaussian variable with zero mean and unit variance. $U[0,2\pi)$ indicates the uniform distribution with the range from 0 to 2$\pi$.

\section{System Model and Problem Formulation}

We consider the uplink of an IRS-assisted mmWave multi-antenna system, where the IRS consists of $N$ passive reflecting antenna elements, i.e., $N = N_{\rm 1} \times N_{\rm 2}$ as a planar array, and the BS is equipped with a UPA of $M = M_{\rm 1} \times M_{\rm 2}$ antennas driven by $L$ RF chains ($L < M$) serving a single-antenna user. For channel estimation, pilots are transmitted by the user, then they are firstly reflected by the IRS before received by the BS. The direct channel component between the user and the BS is not considered due to severe blocked propagation conditions, as commonly adopted in the literature, e.g. \cite{ZCascaded}.


The channel between the user and the IRS and the channel between the IRS and the BS are denoted by $\mathbf{g}\in \mathbb{C}^{N}$ and $\mathbf{G}\in \mathbb{C}^{M \times N}$, respectively. Considering that the antenna arrays at the BS and the IRS are UPAs with standard antenna half-wavelength spacing, the corresponding channels can be expressed as \cite{alkhateeb2014channel}
\begin{equation}
\mathbf{g}=\sum_{k=1}^{N_{\rm p1}}\alpha_k \mathbf{a}_{\rm I}(u_{{\rm I}k}, v_{{\rm I}k}),
\mathbf{G}=\sum_{k=1}^{N_{\rm p2}}\gamma_k \mathbf{a}_{\rm R}(u_{{\rm R}k}, v_{{\rm R}k})\mathbf{a}_{\rm I}^{\rm H}(u_{{\rm I}k}', v_{{\rm I}k}'),
\end{equation}
respectively, where $\alpha_k \in \mathbb{C}$ and $\gamma_k \in \mathbb{C}$ are the corresponding channel gains of the $k$th path, $N_{\rm p1}$ and $N_{\rm p2}$ are the numbers of paths of the corresponding channels, $\mathbf{a}_{\rm I}(u_{{\rm I}k}, v_{{\rm I}k})$, $\mathbf{a}_{\rm R}(u_{{\rm R}k}, v_{{\rm R}k})$, and $\mathbf{a}_{\rm I}(u_{{\rm I}k}', v_{{\rm I}k}')$ are the antenna array response vectors, as elaborated in Appendix A.

At the IRS, we define a diagonal matrix $\mathbf{\Phi}=\mathrm{diag}(\beta_1\mathrm{e}^{j\theta_1},\cdots,\beta_N\mathrm{e}^{j\theta_N})$ as the signal reflection matrix adopted at the IRS, $0\leq \beta_i \leq 1$, is the amplitude coefficient of the $i$th reflecting element of the IRS and $\theta_i\in (0,2\pi]$ is the phase coefficient, $\forall i\in\{1,2,\cdots, N\}$. Note that the reflection matrix can be pre-trained exploiting some coarse beam pre-training techniques, e.g., the synchronization reference signals in current 5G networks.

Let $T$ be the number of channel uses for pilot transmission within one coherence time. Usually, $T$ can be chosen as $\lceil MN/L \rceil$ for a full-rank channel estimation \cite{YuWangWCL}. Let $\mathrm{s}(t)\in\mathbb{C}$ be the pilot symbol with a normalized power satisfying $\mathbb{E}\left\{\mathrm{s}(t)\mathrm{s}^{\rm H}(t)\right\}=1$. Assume that the block-fading channel $\mathbf{G}$ and $\mathbf{g}$ remain unchanged during $T$ channel uses within each coherence time. Then, at time $t$, the corresponding received pilot signal at the BS is
\begin{equation}
\mathbf{r}(t)= \mathbf{G}\mathbf{\Phi}\mathbf{g}\mathrm{s}(t)+\mathbf{n}(t),
\end{equation}
where $\mathbf{n}(t)$ is the additive white Gaussian noise (AWGN) following $\mathcal{CN}(\mathbf{0},\sigma^2_{\rm n}\mathbf{I}_{M})$. The received signal is firstly processed by an analog combiner $\mathbf{W}_{{\rm A}t}\in \mathbb{C}^{M \times L}$, which only imposes phase shifts on the input signal and then the processed signal passes through $L$ low-resolution ADCs. To complete the channel estimation, a subsequent linear digital estimator $\mathbf{W}_{{\rm D}t}\in \mathbb{C}^{M \times L}$ is used. Then, the channel estimate can be expressed as
\begin{equation}\label{eq:h}
\hat{\mathbf{h}}(\mathbf{\Phi})=\mathbf{W}_{\rm D}\mathbf{y},
\end{equation}
where $\mathbf{W}_{\rm D}\!\triangleq\! [\mathbf{W}_{{\rm D}1}, \cdots, \mathbf{W}_{{\rm D}T}]^{\rm T}$, $\mathbf{y}\triangleq [\mathbf{y}(1),\cdots,\mathbf{y}(T)]^{\rm T}$, $\mathbf{y}(t)\triangleq \mathcal{Q}\big(\mathbf{W}_{{\rm A}t}^{\rm H}\mathbf{r}(t)\big)$, $|[\mathbf{W}_{{\rm A}t}]_{ij}|=1$, and $\mathcal{Q}(\cdot)$ represents the operation caused by the quantization of ADCs.

Since the input vector is Gaussian, to improve the tractability of the mathematical problem, we apply the linear model of ADC quantization characterized by the Bussang theorem in \cite{bussgang1952crosscorrelation} which yields:
\begin{align}\label{eq:yt}
\mathbf{y}(t)&\!=\!(1\!-\!\eta_{\rm b})\big(\mathbf{W}_{{\rm A}t}^{\rm H}\mathbf{G}\mathbf{\Phi}\mathbf{g}\mathrm{s}(t)\big)\!+\!(1\!-\!\eta_{\rm b})\mathbf{W}_{{\rm A}t}^{\rm H}\mathbf{n}(t) +\mathbf{e}_{\rm q}(t)\nonumber\\
&\triangleq(1\!-\!\eta_{\rm b})\big(\mathbf{W}_{{\rm A}t}^{\rm H}\mathbf{G}\mathbf{\Phi}\mathbf{g}\mathrm{s}(t)\big)\!+(1\!-\!\eta_{\rm b})\mathbf{e}(t)+\mathbf{e}_{\rm q}(t)\nonumber\\
&=(1\!-\!\eta_{\rm b})\big[\underbrace{\left(\mathrm{s}^{\rm H}(t)\!\otimes\!\mathbf{W}_{{\rm A}t}^{\rm H}\right)}_{\boldsymbol{\zeta}(t)}\mathrm{vec}(\mathbf{G}\mathbf{\Phi}\mathbf{g})\big]\!+\!\tilde{\mathbf{e}}(t),
\end{align}
where $\mathbf{e}(t)\triangleq \mathbf{W}_{{\rm A}t}^{\rm H}\mathbf{n}(t)$, $0 < \eta_{\rm b} < 1$ represents the distortion factor of $b$-bit ADCs \cite{JXuADC}, and $\tilde{\mathbf{e}}(t)\triangleq (1\!-\!\eta_{\rm b})\mathbf{e}(t)+\mathbf{e}_{\rm q}(t)$ represents the noise caused by both AWGN and ADC quantization. Then, stacking vectors of all the $T$ channel uses in a coherence time, the estimated channel vector in \eqref{eq:h} is rewritten as
\begin{align}\label{eq:hh}
\hat{\mathbf{h}}(\mathbf{\Phi})
&=\mathbf{W}_{\rm D}\mathbf{y}=\mathbf{W}_{\rm D}\Big((1\!-\!\eta_{\rm b})\big(\boldsymbol{\zeta}\mathbf{G}\mathbf{\Phi}\mathbf{g}\big)\!+\!\tilde{\mathbf{e}}\Big),
\end{align}
where $\boldsymbol{\zeta}=[\boldsymbol{\zeta}(1),\cdots,\boldsymbol{\zeta}(T)]^{\rm T}$ and $\tilde{\mathbf{e}}=[\tilde{\mathbf{e}}(1),\cdots,\tilde{\mathbf{e}}(T)]^{\rm T}$ are the stacked vectors of $\boldsymbol{\zeta}(t)$ and $\tilde{\mathbf{e}}(t)$, respectively.

By observing \eqref{eq:hh}, we find that due to the limited number RF chains, the dimension of observations $\mathbf{y}(t)$ is only $L$ per estimation, which is insufficient to recover $MN$ channel coefficients. Also, it is difficult to estimate individual channels, $\mathbf{G}$ and $\mathbf{g}$, in a separate manner because $\mathbf{G}$ and $\mathbf{g}$ are both coupled with the IRS reflection matrix $\mathbf{\Phi}$ which is, however, not determined before the channel estimation.

\section{Proposed Estimator for IRS Channel}

\subsection{Problem Reformulation }

Considering the channel estimation in \eqref{eq:hh}, it would be common to express the channels in the angular domain for designing both the analog and digital estimators, $\mathbf{W}_{{\rm A}t}$ and $\mathbf{W}_{\rm D}$. Note that this reformulation would facilitate the estimator design when channel sparsity is further considered. By applying the spatial channel deconstructing approach, we decompose the multi-antenna channels $\mathbf{g}$ and $\mathbf{G}$ as
\begin{equation}\label{eq:gv}
\mathbf{g}_{\rm v}=\mathbf{A}_{\rm i}^{\rm H}\mathbf{g},\quad \mathbf{G}_{\rm v}=\mathbf{A}_{\rm r}^{\rm H}\mathbf{G}\mathbf{A}_{\rm i},
\end{equation}
respectively, where
\begin{align}\label{eq:At}
\mathbf{A}_{\rm i}=\big[&\mathbf{a}_{\rm I}(\tilde{u}_{{\rm I},1}, \tilde{v}_{{\rm I},1}), \cdots, \mathbf{a}_{\rm I}(\tilde{u}_{{\rm I},N_{1}}, \tilde{v}_{{\rm I},1}),\mathbf{a}_{\rm I}(\tilde{u}_{{\rm I},1}, \tilde{v}_{{\rm I},2}), \cdots,\nonumber\\
 &\mathbf{a}_{\rm I}(\tilde{u}_{{\rm I},N_{1}}, \tilde{v}_{{\rm I},N_{2}})\big]\in\mathbb{C}^{N\times N},\nonumber\\
\mathbf{A}_{\rm r}=\big[&\mathbf{a}_{\rm R}(\tilde{u}_{{\rm R},1}, \tilde{v}_{{\rm R},1}), \cdots, \mathbf{a}_{\rm R}(\tilde{u}_{{\rm R},M_{\rm 1}}, \tilde{v}_{{\rm R},1}),\mathbf{a}_{\rm R}(\tilde{u}_{{\rm R},1}, \tilde{v}_{{\rm R},2}),\nonumber\\
&\cdots,\mathbf{a}_{\rm R}(\tilde{u}_{{\rm R},M_{\rm 1}}, \tilde{v}_{{\rm R},M_{\rm 2}})\big]\in\mathbb{C}^{M\times M},
\end{align}
and the virtual angular directions for the two UPAs at the BS and the IRS are chosen as
\begin{align}
\tilde{u}_{{\rm I},p}&\triangleq\frac{1}{2N_{1}}\big(2p-N_{1}-1\big), p\in\{1,2,\cdots,N_{1}\},\nonumber\\
\tilde{v}_{{\rm I},q}&\triangleq\frac{1}{2N_{2}}\big(2q-N_{2}-1\big), q\in\{1,2,\cdots,N_{2}\},\nonumber\\
\tilde{u}_{{\rm R},i}&\triangleq\frac{1}{2M_{1}}\big(2i-M_{1}-1\big), i\in\{1,2,\cdots,M_{\rm 1}\},\nonumber\\
\tilde{v}_{{\rm R},j}&\triangleq\frac{1}{2M_{2}}\big(2j-M_{2}-1\big), j\in\{1,2,\cdots,M_{\rm 2}\},
\end{align}
where $N_{1}, N_{2}$ are the numbers of antennas in the horizontal and vertical direction of the IRS, respectively. $M_{1}, M_{2}$, are defined similarly at the BS.

Then, by substituting \eqref{eq:gv}, the received signal in \eqref{eq:yt} becomes
\begin{align}\label{eq:ytt}
\mathbf{y}(t)
&\!=\!(1\!-\!\eta_{\rm b})\big[\mathbf{W}_{{\rm A}t}^{\rm H}\mathbf{G}\mathrm{diag}\{\mathbf{g}\}\boldsymbol{\phi}\mathrm{s}(t)\big]\!+\!\tilde{\mathbf{e}}(t)\nonumber\\
&\!=\!(1\!-\!\eta_{\rm b})\big[\mathbf{W}_{{\rm A}t}^{\rm H}\mathbf{A}_{\rm r}(\mathbf{G}_{\rm v}\mathbf{g}_{\rm v}\mathbf{1}^{\rm T})\boldsymbol{\phi}\mathrm{s}(t)\big]\!+\!\tilde{\mathbf{e}}(t)\nonumber\\
&\!=\!(1\!-\!\eta_{\rm b})\underbrace{\left(\big[\big(\mathrm{s}^{\rm T}(t)\boldsymbol{\phi}^{\rm T}\big)\!\otimes\!\big(\mathbf{W}_{{\rm A}t}^{\rm H}\mathbf{A}_{\rm r}\big)\big]\right)}_{\mathbf{\Psi}(t)}\underbrace{\left({\rm vec}\big(\mathbf{H}_{\rm v}^{\rm e}\big)\right)}_{\mathbf{h}_{\rm v}}\!+\!\tilde{\mathbf{e}}(t),
\end{align}
where $\boldsymbol{\phi}\triangleq[\beta_1\mathrm{e}^{j\theta_1},\cdots,\beta_N\mathrm{e}^{j\theta_N}]^{\mathrm{T}}\in \mathbb{C}^{N \times 1}$, i.e., $\mathbf{\Phi}={\rm diag}\{\boldsymbol{\phi}\}$, $\mathbf{1}\in\mathbb{R}^{N}$, and $\mathbf{H}_{\rm v}^{\rm e}\!\triangleq\!\mathbf{G}_{\rm v}\mathbf{g}_{\rm v}\mathbf{1}^{\rm T}$. Note that $\boldsymbol{\phi}$ can be any fixed feasible phase shifts in \eqref{eq:ytt} and the choice of $\boldsymbol{\phi}$ does not change the proposed estimator in the following.

Then we stack vectors $\mathbf{y}(t)$ of all the $T$ channel uses together within a coherence time as
\begin{equation}\label{eq:y}
\mathbf{y}=[\mathbf{y}(1), \cdots, \mathbf{y}(T)]^{\rm T}=(1\!-\!\eta_{\rm b})\mathbf{\Psi}\mathbf{h}_{\rm v}+\tilde{\mathbf{e}},
\end{equation}
where $\mathbf{\Psi}\triangleq[\mathbf{\Psi}(1),\cdots,\mathbf{\Psi}(T)]^{\rm T}$, and from \eqref{eq:hh}, we get
\begin{equation}\label{eq:hhat}
\hat{\mathbf{h}}_{\rm v}=(1\!-\!\eta_{\rm b})\mathbf{W}_{\rm D}\mathbf{\Psi}\mathbf{h}_{\rm v}+\mathbf{W}_{\rm D}\tilde{\mathbf{e}},
\end{equation}
where $\hat{\mathbf{h}}_{\rm v}$ is now the desired estimation of the equivalent cascaded channel $\mathbf{h}_{\rm v}$.

Due to the deployment of large IRS, the number of variables in $\mathbf{h}_{\rm v}$ could be large. In mmWave channels, there normally existing some sparsity in the angular domain of the channel. We can further express the cascaded channel in a general form by including the case where a \emph{priori} channel sparsity pattern, say $\mathbf{P}$, is known. This general form helps reduce the computational complexity of channel estimation in \eqref{eq:hhat} if channel sparsity presents. Assume that substantial channel coefficients present only in $N_{\rm v}$ non-sparse angular components, i.e., $\pi(1), \cdots, \pi(N_{\rm v})$. The sparsity pattern is defined as
\begin{equation}
\mathbf{P}=[\mathbf{e}_{\pi(1)}, \mathbf{e}_{\pi(2)}, \cdots, \mathbf{e}_{\pi(N_{\rm v})}\big],
\end{equation}
where $\mathbf{e}_{\pi(i)}$ is a unit vector with the $\pi(i)$th element being 1 and zeros elsewhere and $\mathbf{P}=\mathbf{I}_{MN}$ represents the case where no sparsity is known or presents. Then, the channel coefficients to be estimated in $\mathbf{h}_{\rm v}$ can be rewritten as
\begin{align}\label{eq:hNZ}
\mathbf{h}_{\rm v}^{\rm s}\!=\!\mathbf{P}^{\rm T}\mathbf{h}_{\rm v}\!\triangleq\!\big[\mathbf{e}_{\pi(1)}, \mathbf{e}_{\pi(2)}, \cdots, \mathbf{e}_{\pi(N_{\rm v})}\big]^{\rm T}\mathbf{h}_{\rm v}.
\end{align}
Substituting \eqref{eq:hNZ} into \eqref{eq:hhat}, we have
\begin{equation}\label{eq:hNZe}
\hat{\mathbf{h}}_{\rm v}^{\rm s}\triangleq \mathbf{W}_{\rm D}\big[(1\!-\!\eta_{\rm b})\mathbf{\Xi}\mathbf{h}_{\rm v}^{\rm s}\!+\!\tilde{\mathbf{e}}\big],
\end{equation}
where $\mathbf{\Xi}\triangleq \mathbf{\Psi}\mathbf{P}$. Exploiting the typical estimation criterion as MMSE for continuous variables, we are ready to formulate the channel estimation problem as:
\begin{align}\label{eq:P}
{\rm arg} &\min_{\mathbf{W}_{\rm D},\mathbf{W}_{{\rm A}t}} \mathbb{E}\Big\{\|\hat{\mathbf{h}}_{\rm v}^{\rm s}\!-\!\mathbf{h}_{\rm v}^{\rm s}\|^2_2\Big\},\\\nonumber
&{\rm s.t.}\quad\quad \eqref{eq:hNZ}, \eqref{eq:hNZe}.
\end{align}

To solve the problem in \eqref{eq:P}, we need to optimize $\mathbf{W}_{{\rm A}t}$ and $\mathbf{W}_{\rm D}$. According to \eqref{eq:h}, $\mathbf{W}_{{\rm A}t}$ is restricted as a unity-magnitude valued matrix. Since it is infeasible to apply isotropic pilot directions via the analog hardware of hybrid architecture, which corresponds to independent and identically distributed (i.i.d.) Gaussian $\mathbf{W}_{{\rm A}t}$, we draw phases uniformly from $[0, 2\pi)$ for the construction of $\mathbf{W}_{{\rm A}t}$. Then, we have $[\mathbf{W}_{{\rm A}t}]_{ij}=\frac{1}{\sqrt{M}}e^{j\psi_{ij}}$ with $\psi_{ij}\sim U[0,2\pi)$. Note that the channel estimation should not be directive if no \emph{prior}i channel statistic direction is available. Therefore, this design of analog estimator is able to achieve a uniform performance for an arbitrary channel \cite{YuWangWCL}.

\vspace{-0.3cm}
\subsection{Optimal Linear Digital Estimator }
\vspace{-0.1cm}

The channel estimation problem remains to design the digital estimator $\mathbf{W}_{\rm D}$ by minimizing the MSE between the estimated cascaded channel and the actual one. From \eqref{eq:P}, the MSE is formulated as
\begin{align}\label{eq:MSE}
{\rm MSE}
&=\mathbb{E}\Big\{\|\hat{\mathbf{h}}_{\rm v}^{\rm s}\!-\!\mathbf{h}_{\rm v}^{\rm s}\|^2_2\Big\}\nonumber\\
&=\mathbb{E}\Big\{\|\big[(1\!-\!\eta_{\rm b})\mathbf{W}_{\rm D}\mathbf{\Xi}-\mathbf{I}_{N_{\rm v}}\big]\mathbf{h}_{\rm v}^{\rm s}+\mathbf{W}_{\rm D}\tilde{\mathbf{e}}\|^2_2\Big\}\nonumber\\
&=\sigma^2_{\rm h}(1\!-\!\eta_b)^2{\rm tr}\{\mathbf{W}_{\rm D}\mathbf{\Xi}\mathbf{\Xi}^{\rm H}\mathbf{W}_{\rm D}^{\rm H}\}-\sigma^2_{\rm h}(1\!-\!\eta_b){\rm tr}\{\mathbf{W}_{\rm D}\mathbf{\Xi}\nonumber\\
&\quad +\mathbf{W}_{\rm D}^{\rm H}\mathbf{\Xi}^{\rm H}\}+\sigma^2_{\rm h}N_{\rm v}+\sigma^2_{\tilde{{\rm e}}}{\rm tr}(\mathbf{W}_{\rm D}^{\rm H}\mathbf{W}_{\rm D}).
\end{align}

Applying the law of large numbers, for large $M$, we have
\begin{equation}\label{eq:EWAt}
\mathbb{E}\{\mathbf{W}^{\rm H}_{{\rm A}t}\mathbf{W}_{{\rm A}t}\}\xrightarrow{\rm a.s.}\mathbf{I}_{L},\\
\end{equation}
and assume that $\mathbf{h}_{\rm v}$ satisfies $\mathbb{E}\{\mathbf{h}_{\rm v}\mathbf{h}^{\rm H}_{\rm v}\}=\sigma^2_{\rm h}\mathbf{I}_{NM}$ with $\sigma^2_{\rm h}$ known in advance. Besides, as defined before, $\mathbf{e}(t)\triangleq \mathbf{W}_{{\rm A}t}^{\rm H}\mathbf{n}(t)$ and $\mathbf{n}(t)\sim\mathcal{CN}(\mathbf{0},\sigma^2_{\rm n}\mathbf{I}_{M})$, we can also obtain
\begin{equation}\label{eq:ee}
\mathbb{E}\{\mathbf{e}\mathbf{e}^{\rm H}\}=\sigma^2_{\rm n}\mathbf{I}_{TL},
\end{equation}
where $\sigma^2_{\rm n}$ are known in advance. According to the MMSE criteria, the main task to minimize \eqref{eq:MSE} is to cope with the complex calculation of $\sigma^2_{\tilde{{\rm e}}}$, which is to calculate $\mathbb{E}\big\{\tilde{\mathbf{e}}{\tilde{\mathbf{e}}}^{\rm H}\big\}$:
\begin{align}\label{eq:Eee}
\mathbb{E}\big\{\tilde{\mathbf{e}}{\tilde{\mathbf{e}}}^{\rm H}\big\}\overset{(a)}{=}&(1\!-\!\eta_b)[(1\!-\!\eta_b)\mathbb{E}\{\mathbf{e}\mathbf{e}^{\rm H}\}\nonumber\\
&+\eta_b {\rm diag}\big(\mathbb{E}\{(\mathbf{\Psi}\mathbf{h}_{\rm v}\!+\!\mathbf{e})(\mathbf{\Psi}\mathbf{h}_{\rm v}\!+\!\mathbf{e})^{\rm H}\}\big)]\nonumber\\
\overset{(b)}{=}&(1\!-\!\eta_b)\sigma_{\rm n}^2\mathbf{I}_{TL}\!+\!(1\!-\!\eta_b)\eta_b\sigma_{\rm h}^2{\rm diag}\left(\mathbb{E}\{\mathbf{\Psi}\mathbf{\Psi}^{\rm H}\}\right)\nonumber\\
\overset{(c)}{=}&(1\!-\!\eta_b)(\sigma_{\rm n}^2\!+\!\eta_b{\sigma^2_{\rm h}}N)\mathbf{I}_{TL},
\end{align}
where $(a)$ applies \cite[eq.~(30)]{mezghani2012capacity}, $(b)$ applies \eqref{eq:ee}, and $(c)$ exploits the following expectation as
\begin{align}\label{eq:Epsi}
\nonumber
&\mathbb{E}\big\{\mathbf{\Psi}\mathbf{\Psi}^{\rm H}\big\}\\\nonumber
\overset{(d)}{=}&{\rm diag}\Big(\mathbb{E}\{\left({\rm s}^{\rm T}(1)\boldsymbol{\phi}^{\rm T}\otimes\mathbf{W}_{{\rm A}1}^{\rm H}\mathbf{A}_{\rm r}\right)\left(\boldsymbol{\phi}^*{\rm s}^*(1)\otimes\mathbf{W}_{{\rm A}1}\mathbf{A}_{{\rm r}}^{\rm H}\right)\},\\\nonumber
&\cdots,\mathbb{E}\{\left({\rm s}^{\rm T}(T)\boldsymbol{\phi}^{\rm T}\otimes\mathbf{W}_{{\rm A}T}^{\rm H}\mathbf{A}_{\rm r}\right)\left(\boldsymbol{\phi}^*{\rm s}^*(T)\otimes\mathbf{W}_{{\rm A}T}\mathbf{A}_{{\rm r}}^{\rm H}\}\right)\Big)\\\nonumber
=&{\rm diag}\Big(\mathbb{E}\{{\rm s}^{\rm T}(1)\boldsymbol{\phi}^{\rm T}\boldsymbol{\phi}^*{\rm s}^*(1)\otimes\mathbf{W}_{{\rm A}1}^{\rm H}\mathbf{A}_{\rm r}\mathbf{A}_{\rm r}^{\rm H}\mathbf{W}_{{\rm A}1}\},\cdots,\\\nonumber
&\mathbb{E}\{{\rm s}^{\rm T}(T)\boldsymbol{\phi}^{\rm T}\boldsymbol{\phi}^*{\rm s}^*(T)\otimes\mathbf{W}_{{\rm A}T}^{\rm H}\mathbf{A}_{\rm r}\mathbf{A}_{\rm r}^{\rm H}\mathbf{W}_{{\rm A}T}\}\Big)\nonumber\\
\overset{(e)}{=}&{\rm diag}\Big(N\mathbf{I}_{L},\cdots,N\mathbf{I}_{L}\Big)=N\mathbf{I}_{TL},
\end{align}
where $(d)$ uses \eqref{eq:ytt} and $(e)$ follows by the fact
\begin{equation}\label{eq:EvvH}
\mathbb{E}\big\{{{\rm s}^{\rm T}(t)}\boldsymbol{\phi}^{\rm T}\boldsymbol{\phi}^{\ast}{{\rm s}^{\ast}(t)}\big\}\!=\!\sum_{i=1}^{N}{\beta_{i}^{2}}\mathbb{E}\big\{{{\rm s}^{\rm T}(t)}{{\rm s}^{\ast}(t)}\big\}=N,
\end{equation}
where $\beta_{i}=1$ and $\mathbb{E}\{{\rm s}(t){\rm s}^{\rm H}(t)\}=1$.

Observing the MSE in \eqref{eq:MSE}, it is easy to check that
\begin{equation}
\frac{\partial^{2}({\rm MSE})}{\partial\mathbf{W}_{\rm D}^{2}}
=(1-\eta_b)^2\sigma^2_{\rm h}\mathbf{\Xi}\mathbf{\Xi}^{\rm H}+\sigma^2_{\tilde{\mathbf{e}}}\mathbf{I}_{NM}
\end{equation}
is a positive definite matrix, where $0<\eta_{\rm b}<1$ and typical values of $\eta_{\rm b}$ can be found in  \cite[Table.~I]{YuWangWCL}. Hence, the MSE is convex with respect to $\mathbf{W}_{\rm D}$. Now, we can minimize the MSE in \eqref{eq:MSE} by forcing the following derivative to zero:
\begin{align}\label{eq:pMSE}
&\frac{\partial ({\rm MSE})}{\partial\mathbf{W}_{\rm D}}
=\frac{\partial \mathbb{E}\Big\{||\hat{\mathbf{h}}_{\rm v}^{\rm s}\!-\!\mathbf{h}_{\rm v}^{\rm s}||^2_2\Big\}}{\partial\mathbf{W}_{\rm D}}\nonumber\\
&=(1-\eta_b)^2\sigma^2_{\rm h}\mathbf{\Xi}\mathbf{\Xi}^{\rm H}\mathbf{W}_{\rm D}-(1-\eta_b)\sigma^2_{\rm h}\mathbf{\Xi}+\sigma^2_{\tilde{\mathbf{e}}}\mathbf{W}_{\rm D},
\end{align}
where the variance of $\tilde{\mathbf{e}}$ is $\sigma^2_{\tilde{\mathbf{e}}}=\mathbb{E}\big\{{\tilde{\mathbf{e}}}^{\rm H}\tilde{\mathbf{e}}\big\}=TL(1-\eta_b)(\sigma_{\rm n}^2+\eta_b\sigma_{\rm h}^{2}N)$ from \eqref{eq:Eee}. It yields
\begin{equation}\label{eq:Wmmse}
\mathbf{W}_{\rm D}^{*}\!=\!\frac{1}{1\!-\!\eta_b}\left(\mathbf{\Xi}^{\rm H}\mathbf{\Xi}\!+\!\frac{\sigma^2_{\tilde{\mathbf{e}}}}{(1-\eta_b)^2\sigma^2_{\rm h}}\mathbf{I}_{N_{\rm v}}\right)^{-1}\mathbf{\Xi}^{\rm H}.
\end{equation}
Noting that if no \emph{a priori} channel sparsity is exploited, the burden of matrix inversion computation in \eqref{eq:Wmmse} will be extremely tremendous. Moreover, the number of quantization bits of ADCs, i.e. $b$, directly affects the value of equivalent noise $\sigma^2_{\tilde{\mathbf{e}}}$ which further deteriorates the estimation accuracy.

As a result, the estimated equivalent channel vector in the angular domain is $\hat{\mathbf{h}}_{\rm v}=\big(\mathbf{P}^{\rm T}\big)^{\dagger}\mathbf{W}_{\rm D}^{*}\mathbf{y}$. Then, according to \eqref{eq:ytt}, we can recover the estimated cascaded angular-domain channel by using $\hat{\mathbf{G}}_{\rm v}\hat{\mathbf{g}}_{\rm v}\mathbf{1}^{\rm T}={\rm mat}(\hat{\mathbf{h}}_{\rm v})$. By further multiplying by the transformation matrix $\mathbf{A}_{\rm r}$, we obtain the desired channel estimate\footnote{Note that in time division duplex (TDD) systems, the uplink channel is estimated and the channel can be used for downlink beamforming design even if hardware impairments exist. For this use case, calibration techniques are needed to ensure the reciprocity between the two channels.} as
\begin{align}
\hat{\mathbf{G}}{\rm diag}\{\hat{\mathbf{g}}\}=\mathbf{A}_{\rm r}{\rm mat}(\hat{\mathbf{h}}_{\rm v}).
\end{align}

\vspace{-0.3cm}
\section{Simulation Results}
\vspace{-0.1cm}

%

\begin{figure}
\centering
\includegraphics[width=4in]{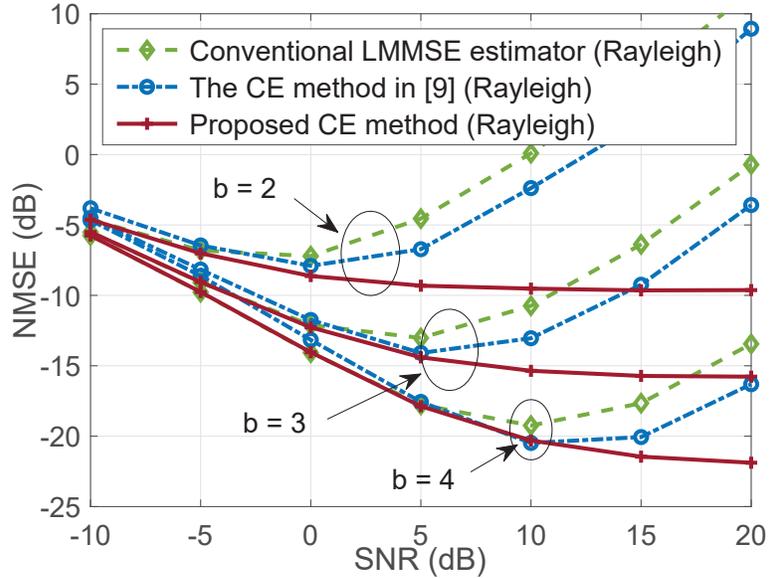}
\caption{NMSE for Rayleigh channels.}
\label{Fig:NMSE1}
\end{figure}

\begin{figure}
\centering
\includegraphics[width=4in]{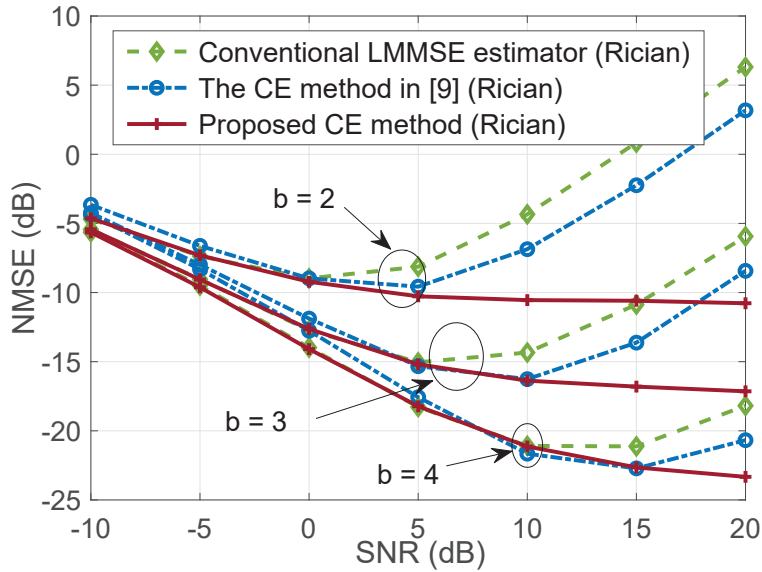}
\caption{NMSE for Rician channels.}
\label{Fig:NMSE2}
\end{figure}


For simulation, we set $N=16$, i.e., $N_{\rm 1}=N_{\rm 2} = 4$, and $M=16$ composed of an $4 \times 4$ UPA driven by $L = 4$ RF chains. As for the IRS reflection matrix adopted in the simulation, the phase coefficient $\theta_i$ is drawn uniformly from [0, 2$\pi$) and $\beta_i$ is normalized to 1, $\forall i\in\{1,2,\cdots, N\}$. We evaluate the performance in terms of the normalized mean squared error (NMSE) of the cascaded channel with a normalized pilot power, i.e., $P_{\rm w}=1$, and the signal-to-noise ratio (SNR) is defined as $10{\rm log}_{10}(P_{\rm w}/\sigma^2_{\rm n})$, which is defined the same between the user-IRS and IRS-BS links.

Fig. 1 compares our proposed channel estimation (CE) method, the conventional LMMSE, and the CE method proposed in \cite{QIRSMISO} of the Rayleigh fading channel \cite{alkhateeb2014channel}, \cite{TS}. It can be seen that when the number of ADCs quantization bits $b$ increases, the performance of all three CE methods improves. In particular, our proposed CE method demonstrates its advantage by effectively suppressing the influence of nonlinear quantization noise and the associated negative effects enhanced by the IRS, while the performance of other two methods deteriorates at high SNRs. Moreover, the estimation error in terms of NMSE eventually saturates for an increasing SNR which is due to the effects of non-vanishing quantization noise. Similar observations can also be found in Fig. 2 under the Rician fading channel with a Rician factor of 10 dB.


Fig. 3 evaluates the effect of the number of antennas on the performance of the proposed CE method. It shows that the performance of the proposed method improves with the increasing number of antennas, while for the other two baseline methods, their performance basically either remain unchanged or become even worse, due to the lack of quantization noise suppression. Fig. \ref{Fig:NMSEsparsity} shows the comparison between our proposed CE method utilizing the sparsity information by the OMP algorithm and our CE method without sparsity at SNR = 10 dB. The sparsity here is defined as the percentage of the number of non-zero channel coefficients divided by number of zero ones in the angular domain. It is obvious that the proposed CE method performs significantly better by incorporating the sparsity information with the burden of matrix inversion computation in \eqref{eq:Wmmse} being reduced. Specifically, after exploiting the sparsity of the channel by the proposed CE method, we can accurately locate non-zero channel coefficients via receive beamforming which facilitates a better coherent combining of received energy. It is demonstrated in Fig. \ref{Fig:NMSEsparsity} that NMSE performs worse with more non-zero channel coefficients of the cascaded channel, as a large portion of signal energy is dissipated during signal propagation.

\begin{figure}[t]
\centering
\includegraphics[width = 4in]{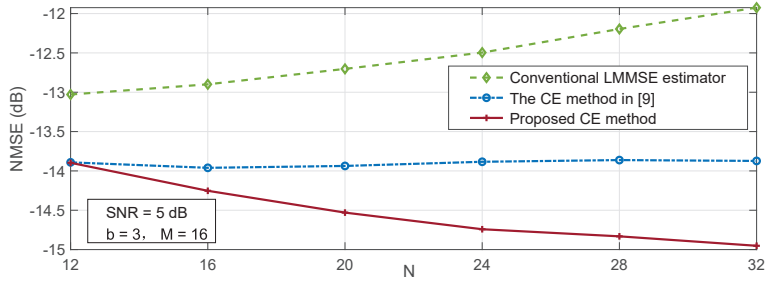}
\caption{The NMSE versus the number of antennas.}
\label{Fig:NMSEantenna}
\end{figure}

\begin{figure}[t]
\centering
\includegraphics[width = 4in]{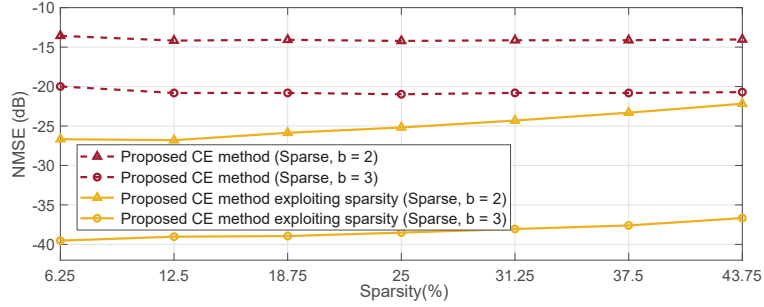}
\caption{The NMSE versus channel sparsity for sparse channels.}
\label{Fig:NMSEsparsity}
\end{figure}

\vspace{-0.3cm}
\section{Conclusion}
\vspace{-0.1cm}

In this paper, an optimized channel estimator was proposed in a closed-form for IRS-assisted multi-antenna systems exploiting hybrid architecture transceivers with low-precision ADCs. The proposed CE method can obtain more accurate cascaded channel estimation with less complexity. For further study, our work can be extended to wideband systems with hardware imperfection of user equipments.

\vspace{-0.3cm}
\appendices
\section{Definition of UPA array response vectors}
\vspace{-0.1cm}

Here, we elaborate the definition of antenna array response vectors $\mathbf{a}_{\rm I}(u_{{\rm I}k}, v_{{\rm I}k})$ , $\mathbf{a}_{\rm R}(u_{{\rm R}k}, v_{{\rm R}k})$, and $\mathbf{a}_{\rm I}(u_{{\rm I}k}', v_{{\rm I}k}')$ for UPA, which mainly follows the definition for uniform linear array (ULA) in \cite{Array3D}.

\begin{figure}[t]
\centering
\includegraphics[width = 4in]{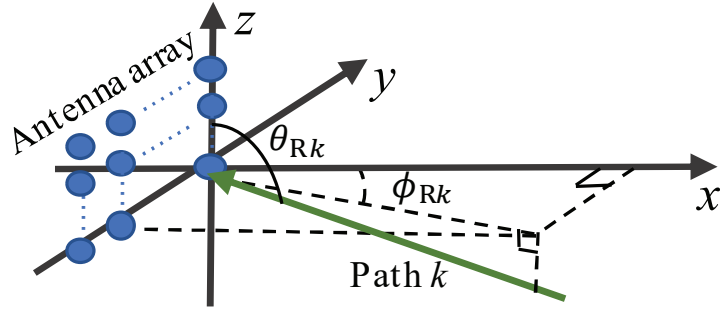}
\caption{An uniform planar array system model schematic diagram.}
\label{Fig:UPA system model}
\end{figure}

Let us take the receiving antenna array at the BS in Fig. \ref{Fig:UPA system model} as an example. The elevation angle and azimuth angle-of-arrival (AOA) of path $k$ are denoted by $\theta_{{\rm R}k}$ and $\phi_{{\rm R}k}$, respectively.

We define two AOA related variables with a carrier wavelength, $\lambda$, and antenna spacing, $d~(d\geq\frac{\lambda}{2})$, as follows
\begin{equation}\label{eq:uv}
u_{{\rm R}k}=\frac{d}{\lambda}{\rm cos}\theta_{{\rm R}k},\quad v_{{\rm R}k}=\frac{d}{\lambda}{\rm sin}\theta_{{\rm R}k}{\rm cos}\phi_{{\rm R}k}.
\end{equation}
Using \eqref{eq:uv}, we define the steering matrix (or called array manifold) $\mathbf{A}_{\rm R}(u_{{\rm R}k}, v_{{\rm R}k})\in \mathbb{C}^{M_{\rm 1} \times M_{\rm 2}}$ as
\begin{equation}
\frac{1}{\sqrt{M}}
 \left[
 \begin{array}{ccc}
 1 &  \cdots & \mathrm{e}^{-j2\pi(M_{\rm 2}-1)v_{{\rm R}k}} \\
 \mathrm{e}^{-j2\pi u_{{\rm R}k}} &  \cdots & \mathrm{e}^{-j2\pi [ u_{{\rm R}k}+(M_{\rm 2}-1)v_{{\rm R}k}]} \\
 \vdots &  \vdots &  \vdots \\
 \mathrm{e}^{-j2\pi(M_{\rm 1}-1)u_{{\rm R}k}} & \cdots & \mathrm{e}^{-j2\pi[(M_{\rm 1}-1)u_{{\rm R}k}+(M_{\rm 2}-1)v_{{\rm R}k}]}
 \end{array}
 \right].
\end{equation}
To simplify the calculation in our paper, the steering matrix $\mathbf{A}_{\rm R}(u_{{\rm R}k}, v_{{\rm R}k})$ is then vectorized as
\begin{equation}
\mathbf{a}_{\rm R}(u_{{\rm R}k}, v_{{\rm R}k})={\rm vec}\big(\mathbf{A}_{\rm R}(u_{{\rm R}k}, v_{{\rm R}k})\big).
\end{equation}

Similarly, we can define the elevation and azimuth AOA of path $k$ at the IRS and angle-of-departure (AOD) of path $k$ at the IRS as $\theta_{{\rm I}k}$ , $\phi_{{\rm I}k}$ and $\theta_{{\rm I}k}'$ , $\phi_{{\rm I}k}'$. The steering matrix $\mathbf{A}_{\rm I}(u_{{\rm I}k}, v_{{\rm I}k})\in \mathbb{C}^{N_{\rm 1} \times N_{\rm 2}}$, $\mathbf{A}_{\rm I}(u_{{\rm I}k}', v_{{\rm I}k}')\in \mathbb{C}^{N_{\rm 1} \times N_{\rm 2}}$ and vector $\mathbf{a}_{\rm I}(u_{{\rm I}k}, v_{{\rm I}k})\in \mathbb{C}^{N_{\rm 1}N_{\rm 2}\times 1}$, $\mathbf{a}_{\rm I}(u_{{\rm I}k}', v_{{\rm I}k}')\in \mathbb{C}^{N_{\rm 1}N_{\rm 2}\times 1}$ can be expressed similarly. It is worth noting that the steering matrix $\mathbf{A}_{\rm I}(u_{{\rm I}k}, v_{{\rm I}k})$ and $\mathbf{A}_{\rm I}(u_{{\rm I}k}', v_{{\rm I}k}')$ are different for different AOAs and AODs at the IRS.

\end{document}